# Single-frame characterization of ultrafast pulses with spatiotemporal orbital angular momentum

Guan Gui,[1] Nathan J. Brooks,[1] Bin Wang,[1] Henry C. Kapteyn,[1,2] Margaret M. Murnane,[1] and Chen-Ting Liao[1,*]

[1]JILA and Department of Physics, University of Colorado and NIST, 440 UCB, Boulder, Colorado 80309, USA
[2]KMLabs Inc., 4775 Walnut Street, Suite 102, Boulder, Colorado 80301, USA
*chenting.liao@colorado.edu

**ABSTRACT:** Light carrying spatiotemporal orbital angular momentum (ST-OAM) makes possible new types of optical vortices arising from transverse OAM. ST-OAM pulses exhibit novel properties during propagation, transmission, refraction, diffraction, and nonlinear conversion, attracting growing experimental and theoretical interest and studies. However, one major challenge is the lack of a simple and straightforward method for characterizing ultrafast ST-OAM pulses. Using spatially-resolved spectral interferometry, we demonstrate a simple, stationary, single-frame method to quantitatively characterize ultrashort light pulses carrying ST-OAM. Using our method, the presence of an ST-OAM pulse, including its main characteristics such as topological charge numbers and OAM helicity, can be identified easily from the unique and unambiguous features directly seen on the raw data—without any need for a full analysis of the data. After processing and reconstructions, other exquisite features, including pulse dispersion and beam divergence, can also be fully characterized. Our fast characterization method allows high-throughput and quick feedback during the generation and optical alignment processes of ST-OAM pulses. It is straightforward to extend our method to single-shot measurement by using a high-speed camera that matches the pulse repetition rate. This new method can help advance the field of spatially and temporally structured light and its applications in advanced metrologies.

1. Introduction

Orbital angular momentum (OAM) of light is a type of intrinsic angular momentum associated with *spatial* spiral phase or wavefront structures of electromagnetic fields that possess a singularity[1,2]. In its standard form, OAM corresponds to a longitudinal phase variation, i.e., averaged angular momentum aligned with the propagation direction of the beam. Since its discovery, this conventional OAM of light has been extensively studied and is the foundation of singular optics[3]. The OAM of light also inspires new physics concepts, such as classical entanglement[4] and self-torque[5], where the former relates to inseparable states of spin angular momentum and OAM (different from direct spin-orbit conversion or coupling[6]), and the latter relates to the recently discovered ability to create pulses with a time-varying OAM. Conventional OAM of light has been employed in applications such as optical trapping, optical tweezers, super-resolution imaging, quantum and classic communication, and scatterometry-based surface metrology[7–11].

Pulsed light can also carry a *transverse* OAM, where the direction of the averaged angular momentum is perpendicular to the averaged wavevector of the beam. Such beams exhibit a swirling phase residing in the spatiotemporal domain, i.e., the $(x,t)$ plane (or equivalent $(x,z)$ plane) in a simplified 2D case, and therefore known as light with spatiotemporal orbital angular momentum (ST-OAM). Light carrying ST-OAM was theoretically predicted[12–14], and first observed as created in the nonlinear filamentation process[15]. Recent research has also shown that ST-OAM pulses can be created in a linear regime – either by a non-cylindrically symmetric 4-f pulse shaper[16,17] or after transmission from a photonic crystal slab with topological response[18]. These studies have inspired further investigations of the fundamental properties of ST-OAM, including its propagation and diffraction[16], reflection and refraction[19], nonlinear conversion[20–22], spin-orbit coupling[23,24], etc. Light with ST-OAM also opens a new way to implement a spatiotemporal differentiator, as recently demonstrated in a symmetry-breaking nanostructure that can generate ST-OAM light or achieve edge enhancement of spatiotemporal imaging[25].

However, despite recent exciting progress in generating beams with ST-OAM, a simple, robust, and comprehensive method to characterize ST-OAM pulses is still lacking. Prior work used a transient-grating supercontinuum spectral interferometer to measure ST-OAM pulses or a spatially resolved scanning interferometer[16,17,20]. Although the transient-grating supercontinuum spectral interferometer is in principle a single-shot method, it requires an extremely complex setup that uses a set of four overlapping pulses and also relies on a third-order nonlinear process and a supercontinuum. Therefore, this approach requires high-intensity ST-OAM pulses that are strong enough to generate nonlinear optical signals while being weak enough to avoid nonlinear propagation inside the optical components[16]. On the other hand, in the case of Mach–Zehnder-like scanning interferometry, measurements of ST-OAM light are restricted to chirped, long-duration pulses due to its limited temporal resolution. This is because scanning interferometry requires a sampling pulse that is much shorter than the ST-OAM pulse as an optical gating pulse to gain resolutions[17,20]. A delay-line is also required in scanning interferometry, making it impractical for characterization with fast feedback. Furthermore, scanning interferometry is insensitive to the global temporal phase of the pulses, although the relative spatiotemporal phase singularity can still be extracted. We note that other spatiotemporal metrologies of laser pulses (e.g., SEA SPIDER[26], SEA TADPOLE[27], STARFISH[28], HAMSTER[29], STRIPED FISH[30], TERMITES[31], INSIGHT[32]), to the best of our knowledge, have not demonstrated the ability to measure complex structured light containing spatiotemporal phase singularities. Among these metrologies, SEA SPIDER relies on nonlinear conversion, which can introduce spatiotemporal distortion into ST-OAM pulses. SEA TADPOLE and STARFISH use optical fibers, which can also reshape ST-OAM pulses. STRIPED FISH, TERMITES, and INSIGHT require iterative algorithms, making them impractical for characterization with fast feedback. HAMSTER builds on an acousto-optic programmable dispersive filter, which can also introduce spatiotemporal distortion on ST-OAM and make its experimental setup more complicated and thus less accessible.

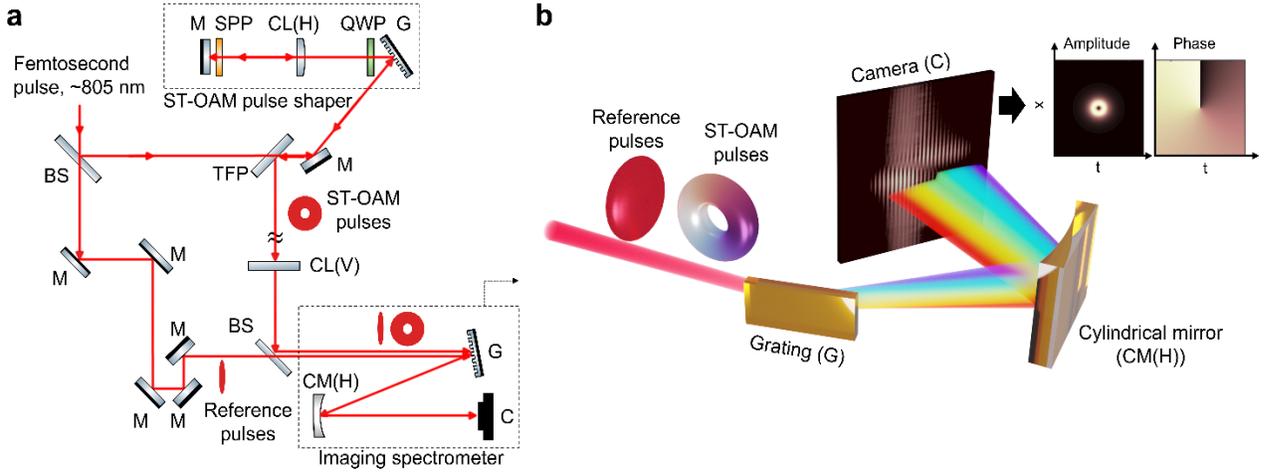

**Figure 1. Experimental setup for the generation and characterization of ST-OAM light.** (a) In our experiment, ST-OAM pulses are generated by a custom pulse shaper and a cylindrical lens (vertically focusing). ST-OAM pulses are characterized by a spatially-resolved spectral interferometer (SRSI), which includes collimated reference pulses and an imaging spectrometer. (b) Schematics of ST-OAM pulse characterization. A typical SRSI raw measurement of reference and ST-OAM pulses of topological charge ℓ = 1 is shown next to the camera. M, mirror; CL(H), cylindrical lens, horizontally focusing; CL(V), cylindrical lens, vertically focusing; CM(H), cylindrical mirror, horizontally focusing; BS, beam splitter; TFP, thin-film polarizer; QWP, quarter-wave plate; SPP, spiral phase plate; G, grating; C, camera.

In this work, we report a simple, stationary, single-frame, single-exposure method, equivalent to a spatially-resolved spectral interferometer (SRSI), to fully characterize ST-OAM pulses, including both Fourier-transform limited and chirped ones. We demonstrate this new quantitative method by measuring and reconstructing ultrashort ST-OAM pulses with different spatiotemporal topological charges, OAM helicities (opposite OAM value; clockwise/counter-clockwise phase swirling direction), spectral dispersions, and spatial divergences. Remarkably, unique and unambiguous features in the raw data can be used to directly identify the presence of ST-OAM pulses, even without a full reconstruction of the pulse. This makes the fast characterization of ST-OAM pulses feasible with very high throughput, as an equivalent ST-OAM mode sorter. In addition, this stationary method can be extended to single-shot, provided that a sufficiently fast camera frame rate that matches ST-OAM pulse repetition rate is used. Moreover, this linear method can be used for weak pulse measurements, potentially down to the single photon level, without requiring a nonlinear optical process. Therefore, our new method promises to greatly simplify the characterization of complex ST-OAM fields for applications in advanced metrology.

2. Method

Our characterization method for ST-OAM pulses is inspired by the measurement of conventional, longitudinal OAM light using a cylindrical lens[33]. When a cylindrical lens focuses a conventional OAM beam with spatial topological charges, multiple lobes can be observed along a diagonal direction at the focal plane due to diffraction of the spiral wavefront of the OAM beam. This unique feature of cylindrical symmetry breaking at the focus, i.e., multiple separated lobes along the diagonal direction, can therefore serve as one of the simplest characterization methods, as an OAM mode sorter, for conventional OAM beams. Since focusing a symmetric beam by a cylindrical lens is equivalent to the 1D Fourier transform of a 2D beam profile in the spatial domain, we can extend this idea and adapt it for ST-OAM light. Considering such use of analogy, our method applies 1D Fourier transform in the temporal domain, shown in the following equation:

$$\widetilde{E_\ell}(x,\omega)e^{i\widetilde{\varphi}_\ell(x,\omega)} = \mathcal{F}_t\{E_\ell(x,t)e^{i\varphi_\ell(x,t)}\}, (1)$$

where $\ell$ is the spatiotemporal topological charge and $\varphi_\ell(x,t)$ is the spatiotemporal phase profile of a cylindrically symmetric, polychromatic beam under the paraxial-wave approximation (collimated beam with sub milli-radian divergence) and the scalar field approximation (linearly polarized light). The image of the beam intensity profile measured at the Fourier plane, i.e., $\tilde{I}(x,\omega) \propto |E_\ell(x,\omega)|^2$, namely, the spatially-resolved spectral intensity, can be characterized directly by an imaging spectrometer. Based on the same analogy, multiple lobes along the diagonal direction should be expected in the spatial-spectral domain $(x,\omega)$. However, because an imaging spectrometer is not capable of detecting spectral phases, such a measurement is insensitive to dispersion presented in ST-OAM pulses, which can be introduced easily by materials that reshape ST-OAM pulses. To acquire the spectral phase for a complete characterization of ST-OAM pulses, we further introduced a Gaussian reference pulse to form an SRSI. Once the reference pulse is well-characterized, for example, using a known Gaussian pulse with a known chirp, both the spatially-resolved spectral intensity and phase can be extracted in a single-frame of a camera, leading to a full reconstruction of ST-OAM pulses in the spatiotemporal $(x,t)$ domain.

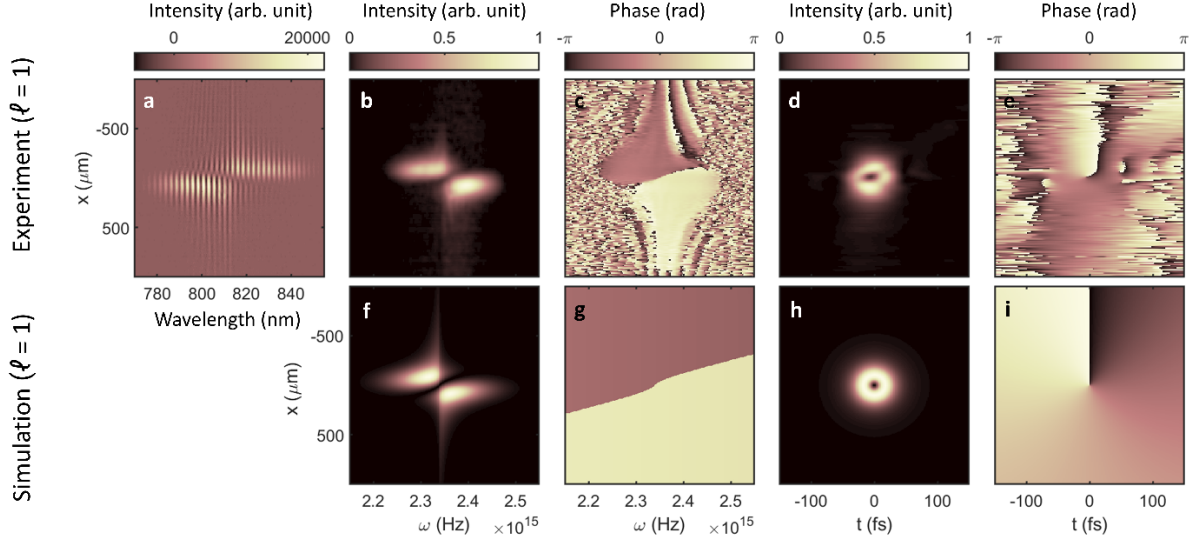

**Figure 2. Experimental measurements, reconstructions, and numerical simulations of ST-OAM light with spatio-temporal topological charge $\ell$ = 1.** (a) Raw experimental measurement by SRSI of ST-OAM pulses ($\ell$ = 1), where the measurement of reference pulse was removed. (b, c) The reconstructed spatial-spectral intensity $\tilde{I}(x,\omega)$ and spatial-spectral phase $\tilde{\varphi}(x,\omega)$, respectively, show two spectral lobes and a π-phase difference between the lobes. (d, e) Spatial-spectral intensity and phase are further used to reconstruct the spatiotemporal intensity $I(x,t)$ and spatiotemporal phase $\varphi(x,t)$ of ST-OAM pulses, respectively. (f-i) Numerical simulations of ST-OAM pulses in the spatial-spectral $(x,\omega)$ domain, f, g, and spatiotemporal $(x,t)$ domain, h,i, agree with the experimental measurements and reconstructions. Top row: experiments and their reconstructions; bottom row: numerical simulations.

Our experimental setup for the generation and characterization of ST-OAM pulses is shown in Fig. 1(a). ST-OAM pulses were generated by a custom non-cylindrically symmetric pulse shaper with pre-compressed pulses from a Ti:sapphire laser oscillator, similar to the one used in Ref.[20]. After the pulse shaper, the ST-OAM beam was *vertically* focused by a cylindrical lens (f=1m) and measured at the focal plane. A Gaussian, collimated reference beam was aligned collinearly with the ST-OAM beam to form an SRSI. After the spectral interferometer, a home-built imaging spectrometer was used to measure the spatially-resolved spectrum. The imaging spectrometer consists of a grating (*vertical* groove direction) and a cylindrical mirror that focuses the beam *horizontally*. A typical spectral interferometric measurement of ST-OAM and reference pulses is also schematically shown in Fig. 1(b). We note that *no scanning* is required in the interferometer – the reference pulse and ST-OAM pulse only need to be set at a *fixed* time delay – that is, a stationary, single-frame or single-shot method once the reference pulse is pre-calibrated.

The spatiotemporal profile of the ST-OAM pulses can be reconstructed from the SRSI measurement. To retrieve the spectral intensity of the ST-OAM pulses, the spatial-spectral $(x,\omega)$ profile of the reference pulse was subtracted from the SRSI measurement, and then a low pass filter was applied numerically to remove the spectral fringes. To retrieve the spectral phase of the ST-OAM pulses, we followed the standard procedures of common spectral interferometry to acquire the spectral phase difference between the ST-OAM and the reference pulse. By measuring the spectral phase of the reference pulse using a typical frequency-resolved optical gating (FROG) setup, the spectral phase of the ST-OAM can be retrieved. Note that there is an ambiguity in the sign of the spectral phase from a FROG measurement when using second-harmonic generation FROG. We identified the sign experimentally by adding a fused silica plate before the FROG, which is known to introduce positive group delay dispersion (GDD) phase. With both spectral intensity and phase, the spatiotemporal profile of ST-OAM pulses can be fully characterized.

### 3. Results

Experimental measurements and reconstructions of ST-OAM light with topological charge $\ell = 1$ are presented in Figs. 2(a-e). A typical SRSI measurement of ST-OAM pulses is shown in Fig. 2(a), where the spatial-spectral profile of the reference pulses was removed. From the single-frame spectral interferometric measurement, a two-lobe structure can be observed along the diagonal direction in the $(x,\lambda)$ domain. This spatial-spectral profile is then used to reconstruct the spatial-spectral intensity, $\tilde{I}(x,\omega)$, as presented in Fig. 2(b). The two-lobe structure along the diagonal direction in the $\tilde{I}(x,\omega)$ plot indicates a spatial chirp in the ST-OAM pulse, i.e., the spectrum from the top ($x<0$) and the bottom part ($x>0$) of the ST-OAM pulse are red- and blue-shifted, respectively, which originates from the spiral phase of the ST-OAM pulse in the spatiotemporal $(x,t)$ domain. The spectral interferometric measurement in Fig. 2(a) also shows some spectral fringes on the two-lobe structure, and these spectral fringes are shifted by a π-phase difference, comparing the top and the bottom portions. Based on these spectral fringes, the spatial-spectral phase, $\tilde{\varphi}(x,\omega)$, can be reconstructed, as presented in Fig. 2(c), showing a π-phase difference between the two spectral lobes. Using the experimentally reconstructed $\tilde{I}(x,\omega)$ and $\tilde{\varphi}(x,\omega)$ from Figs. 2(b,c), the ST-OAM pulse can be retrieved in the spatiotemporal $(x,t)$ domain for $I(x,t)$ and $\varphi(x,t)$, as shown in Figs. 2(d,e). The reconstructed ST-OAM pulse has a donut-shaped and a spiral phase in the $(x,t)$ domain. We also performed numerical simulations to confirm our experimental observations, measurements, and reconstructions. The simulations agree with our measurement of ST-OAM pulses very well, as shown in Figs. 2(f-i) for spatial-spectral intensity, spatial-spectral phase, spatiotemporal intensity, spatiotemporal phase, respectively. More information on the simulations can be found in the Supporting Information. We note that the reconstructed ST-OAM pulse is near the Fourier-transform limits (pulse duration ~45 fs) – such high time resolution is not achievable using past scanning interferometer methods. Moreover, shorter-pulse, few-cycle, ST-OAM pulses could be measured using our SRSI method since achieving a broader spectral range and higher resolving power from a spectrometer is straightforward.

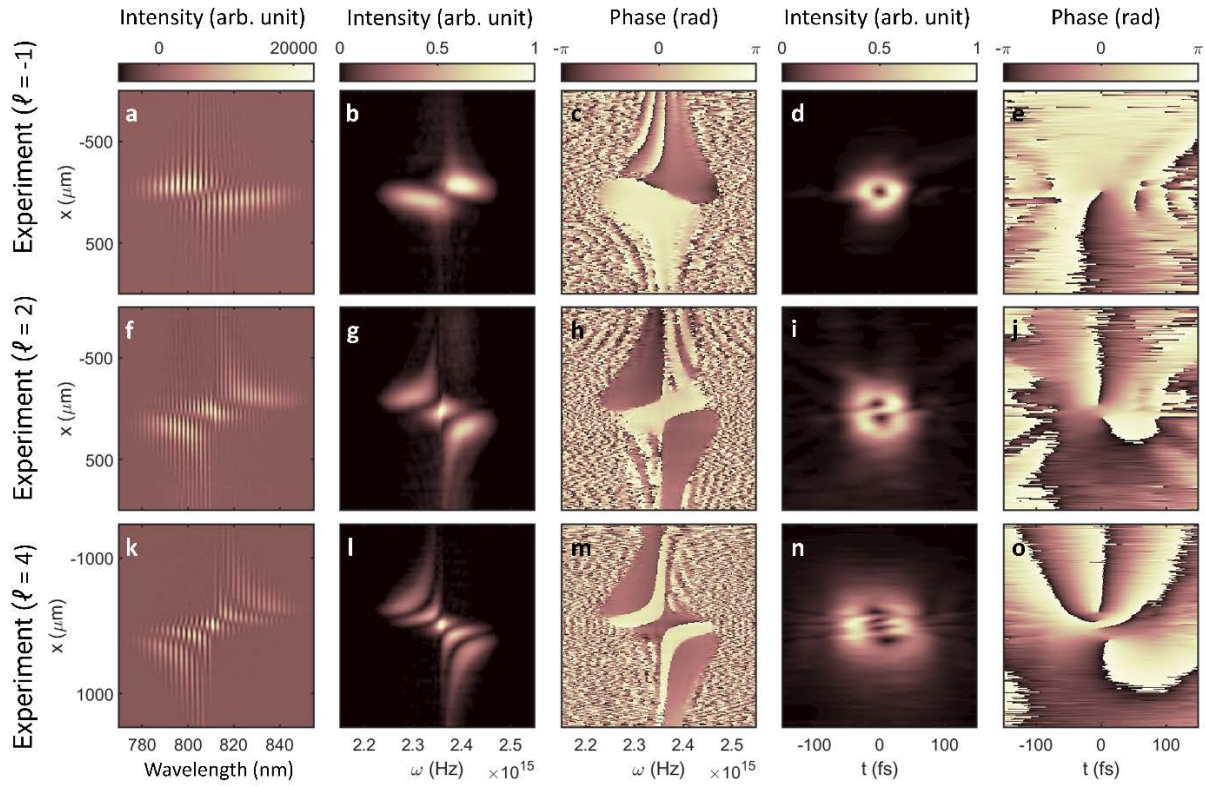

**Figure 3. Experimental measurements and their reconstructions of ST-OAM light with spatiotemporal topological charge $\ell$ = -1, 2, 4.** Spectral interferometric measurements of ST-OAM pulses of charges $\ell$ = -1 (a), $\ell$ = 2 (f), $\ell$ = 4 (k), where the measurement of reference pulse has been removed. The reconstructed spatial-spectral intensities $\tilde{I}(x,\omega)$ of $\ell$ = -1 (b), $\ell$ = 2 (g), $\ell$ = 4 (i) show multiple-lobe structure along diagonal directions, where the number and direction of the lobes indicate the topological charge and helicity of the pulses. The spatial-spectral phases $\tilde{\varphi}(x,\omega)$ (c,h,m) show π-phase steps between each spectral lobe. The reconstructed spatiotemporal intensities $I(x,t)$ (d,i,n) and $\tilde{\varphi}(x,\omega)$ phases (e,j,o) show one or multiple singularities near the center of the beams, resembling a torus of one, two, and four. The split of phase singularities has been predicted and observed by both simulations[13,34] and experiment[17].

To showcase the quantitative capability of this SRSI method, in the following sections and figures, we demonstrate the characterization of ST-OAM pulses with different spatiotemporal topological charges, including $\ell = -1, 2, 4$, as presented in Fig. 3. For pulses with charge $\ell = -1$, their spectral interferometric measurements and reconstructions are shown in Figs. 3(a-e). In the spatial-spectral intensity plot (Fig. 3(a)), a two-lobe structure is clearly visible, while the two lobes are oriented in the *opposite* diagonal direction, compared with that from $\ell = 1$ pulse (Fig. 2(a)). In the spatial-spectral plot (Fig. 3(c)), a π-phase difference is also observed between the two lobes. The spatiotemporal reconstructions of $\ell = -1$ pulses are shown in Figs. 3(d,e), where a donut-shaped pulse with a spiral phase can be seen in the $(x,t)$ domain, although the spiral phase presents a different OAM helicity—increasing phase in the clockwise direction in Fig. 2(e) and decreasing phase in the clockwise direction in Fig. 3(e). Experimental measurements and reconstructions of $\ell = 2$ pulses are shown in Figs. 3(f-j). The spatial-spectral intensity plot, Fig. 3(g), shows a three-lobe structure along the diagonal direction, and there are π-phase differences in-between each spectral lobe. The spatiotemporal intensity of $\ell = 2$ pulses, displayed in Fig. 3(i), shows a donut-shaped pulse with two holes close to the center – a case resembling a torus of genus two. The phase profile in Fig. 3(j) shows two phase singularities (holes in the intensity profile), each with a 2π spiral phase, giving rise to a 4π accumulated phase. We note that the split of phase singularities has been predicted by simulations[13,34] and observed by former experiment[17], especially for ST-OAM pulses generated by a 4-f pulse shaper. Experimental measurements and reconstructions of $\ell = 4$ pulses are shown in Figs. 3(k-o). In the spatial-spectral domain, five spectral lobes can be observed, which also have π-phase differences between each other. In the spatiotemporal reconstructions, a pulse resembling a torus of genus four is shown, together with four 2π-phase singularities. Based on our observations, measurements, and simulations of ST-OAM pulses with different charges and helicities (Figs. 2-3), we establish a general rule of thumb: the spatial-spectral intensity plots carry unique and unambiguous signatures of the spatiotemporal topological charge numbers and the OAM helicity. Therefore, ST-OAM pulses are readily identifiable, depending on the number of the spectral lobes and their diagonal directions, respectively. Therefore, the presence of ST-OAM pulses can be directly identified from the unprocessed, spectral interferometric images directly seen on a camera. This direct observation of ST-OAM pulses by an untrained eye prior to a complete reconstruction is crucial, which enables a high throughput metrology tool for ST-OAM pulses. As a result, our single-frame method represents a convenient measurement approach, providing advantages, ease-of-use, high-speed, and enhanced capabilities.

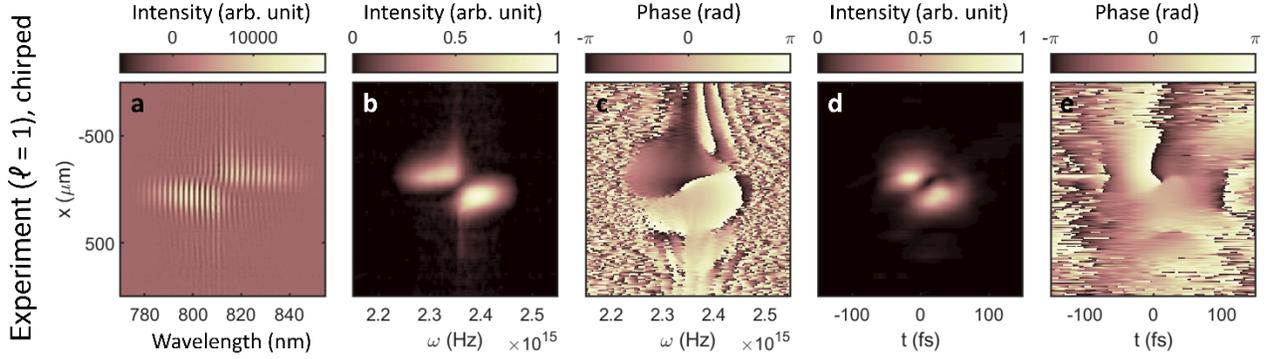

**Figure 4. Experimental measurements and their reconstructions of ST-OAM light with spectral dispersion.** (a) Spectral interferometric measurement of ST-OAM pulses of topological charge $\ell = 1$ passing a 6 mm fused silica plate, where the measurement of reference pulse has been removed. The reconstructed spatial spectral intensity $\tilde{I}(x,\omega)$ (b) is almost the same as unchirped pulses, while the spatial spectral phase $\tilde{\varphi}(x,\omega)$ (c) shows a quadratic spectral phase in addition to the to the π-phase structure. The spatiotemporal reconstructions of the chirped ST-OAM pulse (d,e) form a two-lobe structure in the $(x,t)$ domain instead of a donut shape for the unchirped pulse.

Next, we further investigate how this SRSI method can be used to characterize the dispersion of ST-OAM pulses. Former studies have shown that dispersion can significantly reshape the spatiotemporal profile of ST-OAM pulses[20,35], although the spectral intensity remains unchanged. ST-OAM pulses can acquire dispersion (i.e., chirp) as the beam propagates through optical elements such as lenses, beam splitters, filters, etc. Therefore, it is crucial to characterize ST-OAM pulses with dispersion and distinguish them from unchirped, Fourier-transform limited pulses. In our experiment, we used a 6 mm fused silica plate to introduce $\sim 210$ fs$^2$ of GDD onto $\ell = 1$ ST-OAM pulses. Experimental measurements and reconstructions of chirped ST-OAM pulses are presented in Fig. 4. We can see that the spatial-spectral intensity of the chirped pulses (Fig. 4(b)) is almost identical to the unchirped pulses (Fig. 2(b)), as expected since dispersion does not modify the spectral intensity. However, the spatial-spectral phase of the chirped pulses (Fig. 4(c)), shows a *quadratic spectral* phase, i.e., GDD, in addition to the π-phase structure from the unchirped pulses (Fig. 2(c)). As a result, the spatiotemporal profile of the chirped pulses forms a two-lobe structure in the spatiotemporal domain (Fig. 4(d)) instead of a donut shape from the unchirped pulse (Fig. 2(d)). The reconstructions of chirped ST-OAM pulses are consistent with our simulations presented in the Supporting Information. Both experiment and simulation results confirm that our method clearly distinguishes ST-OAM pulses with different dispersions.

Finally, we investigate and demonstrate how this SRSI method can be used to characterize divergences in ST-OAM pulses. Former studies have explored the propagation and diffraction properties of ST-OAM pulses[16], which suggest that the spatial phase plays an important role in the spatiotemporal profile of the pulses. Here we characterize ST-OAM pulses before and after the focal plane ±0.3 Rayleigh range, where ST-OAM beams are converging or diverging, i.e., carrying negative or positive *quadratic spatial* phases. The experimental measurements and reconstructions of ST-OAM pulses are shown in Fig. 5. The spatial-spectral phase (Figs. 5(c,h)) shows a negative or positive quadratic spatial phase, which leads to different spatiotemporal profiles for convergent or divergent ST-OAM beams (Figs. 5(d,i)). The reconstructions of ST-OAM pulses with different divergences are also confirmed by our simulations, as presented in the Supporting Information. Note that the spatial-spectral phase plays a key role in reconstructing ST-OAM pulses, which can be well-characterized by our single-frame method, either for chirped pules (*spectral* phase) or divergent beams (*spatial* phase).

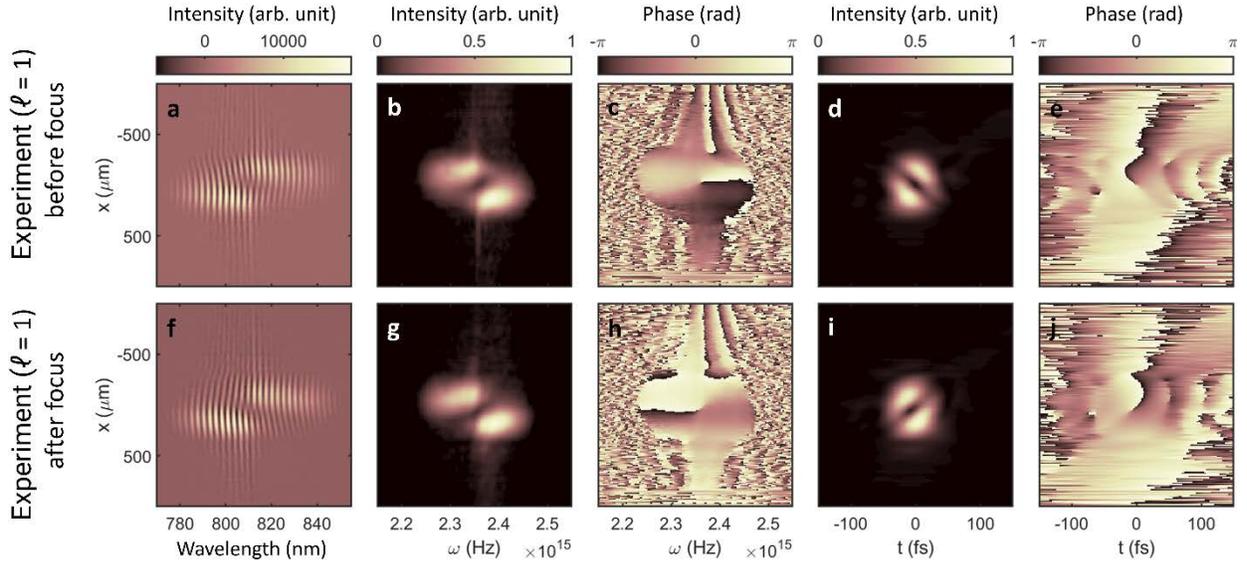

**Figure 5. Experimental measurements and their reconstructions of ST-OAM light with spatial divergence.** Spectral interferometric measurement of ST-OAM pulses of topological charge $\ell$ = 1, at ~0.3 Rayleigh range before (a) and after (f) the focal plane, where the measurement of reference pulse has been removed. The reconstructed spatial spectral intensities $\tilde{I}(x,\omega)$ (b,g) still form a two-lobe structure. The spatial-spectral phases $\tilde{\varphi}(x,\omega)$ show negative (c) or positive (h) quadratic spatial phases in addition to the π-phase structure, which leads to different spatiotemporal profiles for convergent (d,e) or divergent (e,j) ST-OAM beams.

4. Conclusion

In summary, we demonstrate a simple, stationary, single-frame, single-exposure method based on spatially-resolved spectral interferometry to fully characterize ST-OAM pulses with different spatiotemporal topological charges, OAM helicities, spectral dispersions, and spatial divergences. In our new method, the presence of an ST-OAM pulse, including its charge numbers and OAM helicity, can be directly identified from the unique and unambiguous features seen on the raw spectral images taken on a camera, even before any processing and reconstructions. This remarkable result indicates that a simple measurement is sufficient to capture the quantitative signatures of ST-OAM pulses. This characterization method, as an ST-OAM mode sorter, allows high-throughput and quick feedback during the generation and optical alignment of ST-OAM pulses. This method can also characterize other exquisite features such as pulse (spectral) dispersion and beam (spatial) divergence via measuring the spatial-spectral phase after processing and reconstructions. Compared with prior methods for ST-OAM light characterization, our method utilizes a linear experimental procedure without a nonlinear process and thus can be used for low-light measurement. Extending our single-frame method to single-shot measurement is straightforward by using a high-speed camera. Therefore, we expect this new metrology method will substantially benefit fundamental light science research on ST-OAM pulses and future applications.

## ASSOCIATED CONTENT

### Funding Sources

DE-FG02-99ER14982, FA9550-16-1-0121 DGE-1650115.

### Notes

H.K. has a co-affiliation at KMLabs Inc. H.K. and M.M. have a financial interest in KMLabs. Other authors declare no conflicts of interest.

## ACKNOWLEDGMENT

We acknowledge funding from the Department of Energy BES Award No. DE-FG02-99ER14982 for the development of the new characterization technique, and a MURI grant from the Air Force Office of Scientific Research under Award No. FA9550-16-1-0121 for the experimental setup, N.J.B. acknowledges support from the National Science Foundation Graduate Research Fellowships (Grant No. DGE-1650115).

# Single-frame characterization of ultrafast pulses with spatiotemporal orbital angular momentum: supporting information

**ABSTRACT:** This document provides supporting information to "Single-frame characterization of ultrafast pulses with spatiotemporal orbital angular momentum".

**Numerical simulation method.** We performed numerical simulations of light with spatiotemporal orbital angular momentum (ST-OAM) to verify our experimental measurements and reconstructions. Our simulation started with ultrafast pulses with a Gaussian-shaped spectrum ($\lambda_c = 805\ nm, \Delta\lambda_{FWHM} = 40\ nm$) and a Gaussian-shaped spatial profile ($D_{FWHM} = 2.5\ mm$). The custom pulse shaper, a 4-f system as shown in Fig.1 (a), was simulated by Fourier transform between $(x, t)$ and $(x, \omega)$ domains. In the $(x, \omega)$ domain, a spiral phase was added to the ST-OAM pulse to mimic the spiral phase plate (SPP) placed at the Fourier plane in the pulse shaper. After the pulse shaper, a quadratic spatial phase was added to the ST-OAM pulse to simulate the cylindrical lens ($f = 1\ m$, vertically focusing), and then the spatiotemporal profile of ST-OAM pulses around the focal plane was calculated based on free-space propagation with Fresnel approximation. The following sections present our simulation results of ST-OAM pulses with different topological charges, OAM helicities, pulse dispersions, and beam divergences. Note that part of the simulation results is shown in Fig. 2(f-i) of the main article.

**ST-OAM pulses with different charges and OAM helicity.** Simulated ST-OAM pulses with spatiotemporal topological charges $\ell = -1, 2, 4$ are simulated and shown in Fig. S1. The spatial-spectral intensity plots (Fig. S1(a,e,i)) show multiple-lobe structures along different diagonal directions, where the number and direction of spectral lobes indicate the spatiotemporal topological charge number and OAM helicity of ST-OAM pulses. As presented in Fig. S1(b,f,j), the spatial-spectral phases plots show π-phase differences between each spectral lobe. The spatial-spectral simulations, i.e., $\tilde{I}(x, \omega)$ and $\tilde{\varphi}(x, \omega)$, agree with our experimental measurements as shown in Fig. 3. In the $(x, t)$ domain, phase singularities in the simulated ST-OAM pulses are overlapped (Fig. S1(d,h,l)), which leads to donut-shaped pulses with a single hole near the center (Fig. S1(c,g,k)). Although our simplified simulations show major characteristics of ST-OAM light, e.g., multiple spectral lobes, π-phase steps, and phase singularities, it does not reproduce the split of phase singularities. The splitting of phase singularities has been predicted by other numerical simulations[1,2] and observed experimentally[3], especially for ST-OAM pulses generated by a 4-f pulse shaper. Such a split of phase singularities was observed in our experimental measurements, as shown in Fig. 3.

**ST-OAM pulses with dispersion.** Simulation results of ST-OAM pulses with dispersion are shown in Fig. S2. A GDD of $210 fs^2$ was added to ST-OAM pulses with $\ell = 1$ to simulate the dispersion from a 6 mm fused silica used in our experiment. The simulated spatial-spectral intensity, as presented in Fig. S2(a), is identical to the simulation of the unchirped, Fourier-transform limited pulse, as shown in Fig. 2(f), while the simulated spatial-spectral phases show an extra quadratic *spectral* phase in addition to the π-phase structure. The spatiotemporal $(x, t)$ profile of the chirped ST-OAM pulse has a two-lobe structure instead of a donut shape, which agrees with our experimental reconstructions in Fig. 4(d).

**ST-OAM pulses with different divergences.** Simulation results of ST-OAM pulses with different divergences are shown in Fig. S3. ST-OAM pulses with $\ell = 1$ were simulated before and after the focal plane $\pm 0.3$ Rayleigh range, using free-space propagation with Fresnel approximation. In Fig. S3(b,f), the simulated spatial-spectral phases show extra negative or positive quadratic spatial phases on top of the π-phase structure, which corresponds to convergent and divergent beams, respectively. The simulated spatiotemporal profile of convergent or divergent ST-OAM beams is shown in Fig. S3(c,g), which agrees with our experimental measurements in Fig. 5(d,i).

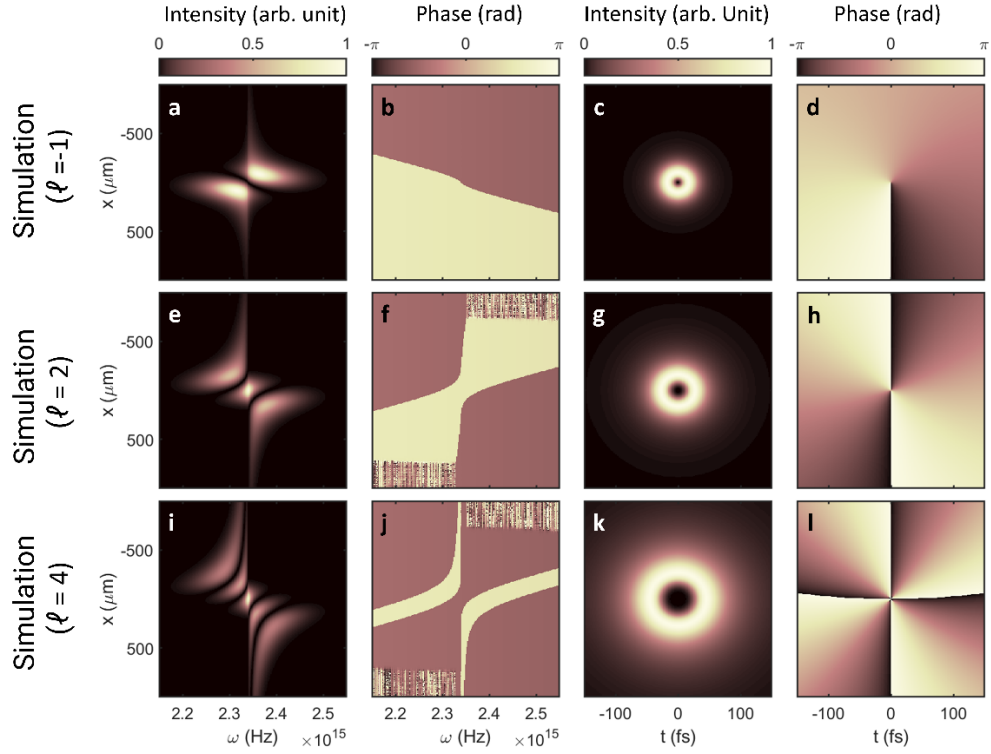

**Figure S1. Simulations of ST-OAM pulses with different spatiotemporal topological charge.** The simulated spatial-spectral intensities $\tilde{I}(x,\omega)$ of $\ell$ = -1 (a), $\ell$ = 2 (e), $\ell$ = 4 (i) show multiple-lobe structure along diagonal directions, where the number and direction of the lobes indicate the topological charge and OAM helicity of the pulses. The spatial-spectral phases $\tilde{\varphi}(x,\omega)$ (b,f,j) show π-phase steps between each spectral lobe. The simulated spatiotemporal intensities $I(x,t)$ (c,g,k) and phases $\varphi(x,t)$ (d,h,l) show one or multiple singularities overlapped near the center of the beams.

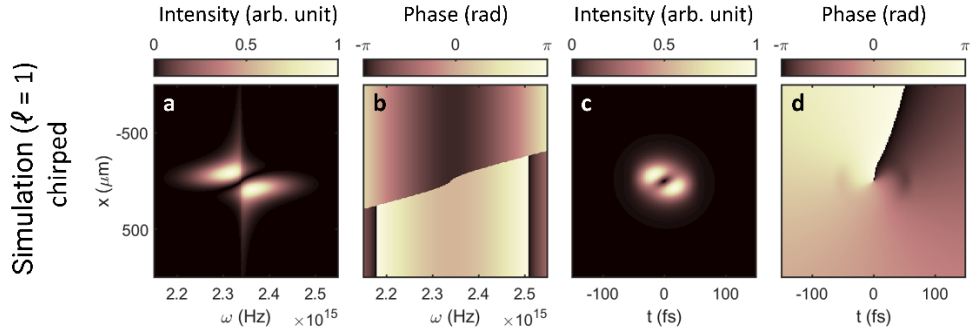

**Figure S2. Simulations of ST-OAM pulses with spectral dispersion.** A GDD of 210 $fs^2$ was introduced to ST-OAM pulses of $\ell$ = 1. The simulated spatial spectral intensity (a) is identical to the unchirped pulse, and spatial spectral phase (b) shows an extra quadratic spectral phase in addition to the π-phase steps. The simulated spatiotemporal intensity (c) and phase (d) are consistent with our experimental reconstructions.

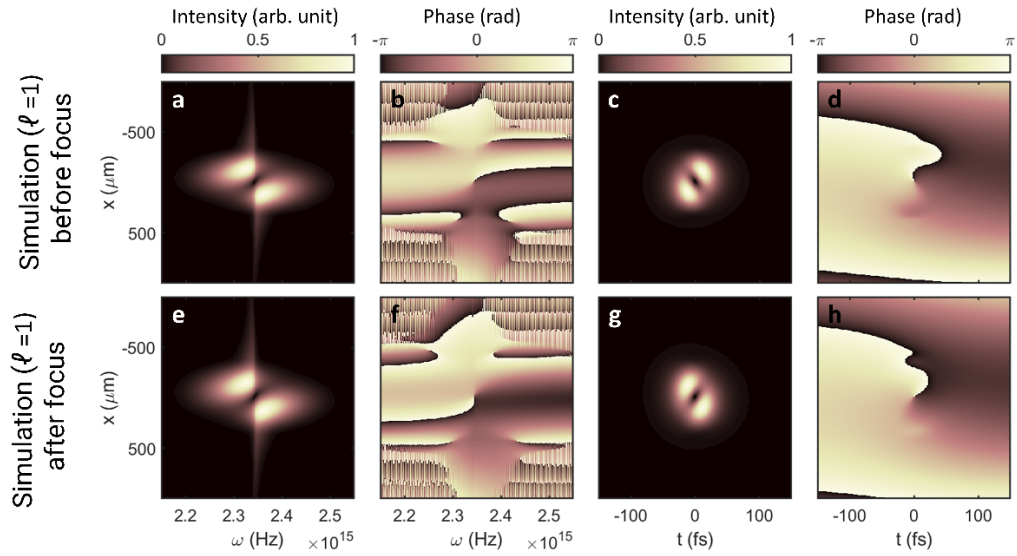

**Figure S3. Simulations of ST-OAM pulses with spatial divergences.** ST-OAM pulses of $\ell$ = 1 were simulated before and after the focal plane $\pm 0.3$ Rayleigh range. The simulated spatial spectral intensities (a,e) still have two spectral lobes, and spatial spectral phase (b,f) shows an extra quadratic spatial phase on top of the $\pi$-phase steps, corresponding to the convergent and divergent cases respectively. The simulated spatiotemporal intensity (c,g) and phase (d,h) consist well with our experimental reconstructions.